\documentclass[preprint]{elsarticle}
\usepackage{xcolor}
\usepackage{multicol}
\usepackage{graphicx}
\usepackage[titlenumbered,linesnumbered,ruled,vlined]{algorithm2e}

\SetKwInput{KwInput}{Input}                
\SetKwInput{KwOutput}{Output}              
\SetKwFor{Dor}{do}{}{endloop}

\usepackage{comment}
\usepackage[%
rm={oldstyle=false,proportional=true},%
sf={oldstyle=false,proportional=true},%
tt={oldstyle=false,proportional=true,variable=true},%
qt=false%
]{cfr-lm}
\usepackage{amsmath}
\usepackage{color}

\usepackage{pdfcomment}

\DeclareMathOperator{\diag}{diag}

\DeclareMathOperator{\trace}{trace}

\newcommand{\set}[1]{\left\{#1\right\}}
\newcommand{\xcap}[1]{X^{(#1)}}
\newcommand{\wcap}[1]{W^{(#1)}}
\newcommand{\pcap}[1]{P^{(#1)}}
\newcommand{\scap}[1]{S^{(#1)}}
\newcommand{\rcap}[1]{R^{(#1)}}
\newcommand{\ccap}[1]{C^{(#1)}}

\newcommand{\xcapt}[1]{X^{(#1)T}}

\newcommand{\scapt}[1]{S^{(#1)T}}

\newcommand{\thcap}[1]{\Theta^{(#1)}}

\makeatletter
\def\verbatim@font{\fontfamily{cmtt}\scriptsize\selectfont}
\makeatother

\DeclareFontFamily{U}{MnSymbolC}{}
\DeclareSymbolFont{MnSyC}{U}{MnSymbolC}{m}{n}
\DeclareFontShape{U}{MnSymbolC}{m}{n}{
    <-6>  MnSymbolC5
   <6-7>  MnSymbolC6
   <7-8>  MnSymbolC7
   <8-9>  MnSymbolC8
   <9-10> MnSymbolC9
  <10-12> MnSymbolC10
  <12->   MnSymbolC12%
}{}
\DeclareMathSymbol{\powerset}{\mathord}{MnSyC}{180}

\usepackage[utf8]{inputenc}
\usepackage{fourier}
\usepackage{array}
\usepackage{booktabs}
\usepackage{tabularx}
\usepackage{makecell}

\newcolumntype{Z}{ >{\centering\arraybackslash}X }

\begin{document}
\title{Accelerating an Iterative Eigensolver for Nuclear Structure Configuration Interaction Calculations on GPUs using OpenACC}
\author[1]{Pieter Maris\corref{cor1}}
\ead{pmaris@iastate.edu}

\author[2]{Chao Yang}
\ead{cyang@lbl.gov}

\author[3]{Dossay Oryspayev}
\ead{doryspaye@bnl.gov}

\author[4]{Brandon Cook}
\ead{bgcook@lbl.gov}

\cortext[cor1]{Corresponding author}
\address[1]{Department Of Physics and Astronomy, Iowa State University, Ames, IA, USA}
\address[2]{Computational Research Division, Lawrence Berkeley National Laboratory, Berkeley, CA, USA}
\address[3]{Computational Science Initiative, Brookhaven National Laboratory, Upton, NY, USA}
\address[4]{National Energy Research Scientific Computing Center, Lawrence Berkeley National Laboratory, Berkeley, CA, USA}

\begin{abstract}
To accelerate the solution of large eigenvalue problems arising from many-body calculations in nuclear physics on distributed-memory parallel systems equipped with general-purpose Graphic Processing Units (GPUs), we modified a previously developed hybrid MPI/OpenMP implementation of an eigensolver written in FORTRAN 90 by using an OpenACC directives based programming model. Such an approach requires making minimal changes to the original code and enables a smooth migration of large-scale nuclear structure simulations from a distributed-memory many-core CPU system to a distributed GPU system.  However, in order to make the OpenACC based eigensolver run efficiently on GPUs, we need to take into account the architectural differences between a many-core CPU and a GPU device. Consequently, the optimal way to insert OpenACC directives may be different from the original way of inserting OpenMP directives.  We point out these differences in the implementation of sparse matrix-matrix multiplications (SpMM), which constitutes the main cost of the eigensolver, as well as other differences in the preconditioning step and dense linear algebra operations. We compare the performance of the OpenACC based implementation executed on multiple GPUs with the performance on distributed-memory many-core CPUs, and demonstrate significant speedup achieved on GPUs compared to the on-node performance of a many-core CPU.  We also show that the overall performance improvement of the eigensolver on multiple GPUs is more modest due to the communication overhead among different MPI ranks.
\end{abstract}


\maketitle              

\section{Introduction\label{sec:intro}}

One of the most challenging problems in the computational study of the structure of atomic nuclei is the solution of a large-scale eigenvalue problem 
\begin{equation}
H \psi = \lambda \psi,
\label{eq:hev}
\end{equation}
where $H$ is an approximation to the many-body Hamiltonian associated with 
a target nucleus, $\lambda$ is an eigenvalue of $H$ and $\psi$ is the corresponding eigenvector~\cite{Barrett:2013nh,Vary2018Ab,Caprio:2019yxh}.  When $H$ is 
approximated in a subspace spanned by Slater determinants of 
some single-particle basis functions, the finite-dimensional eigenvalue
problem defined in \eqref{eq:hev} is often called a Configuration Interaction (CI) approximation, and the matrix $H$ is often referred to as
a CI Hamiltonian.

Due to the many-body nature of the eigenvalue problem, the dimension 
of the CI Hamiltonian $H$, which is a function of the number of nucleons and
a basis truncation parameter, can become extremely large.  
However, $H$ is generally very sparse. Therefore, iterative methods that can take advantage of an efficient implementation of a Sparse Matrix Vector multiplication (SpMV)
procedure are often preferred, especially when only a limited number of eigenpairs of $H$ are needed~\cite{saad:book,sternberg2008}. 

In nuclear physics, one is often interested in a few (five to ten) low-lying eigenpairs of $H$. In recent work \cite{SHAO20181}, we have shown that these eigenvalues can be computed efficiently by using the Locally Optimal Block Preconditioned Conjugate Gradient (LOBPCG) method \cite{Knyazev2001}. The advantages of the LOBPCG algorithm, which we will describe with some detail in the next section, over the widely used Lanczos algorithm \cite{Lanczos1950}, are
\begin{itemize}
\item The algorithm is a block method that allows us to multiply $H$ with several vectors simultaneously.
That is, instead of an SpMV one performs an SpMM of a sparse square matrix on a tall skinny matrix at every iteration, which introduces an additional level of concurrency in the computation and enables us to exploit data locality better.
\item The algorithm allows us to make effective use of approximations to several eigenvectors.
\item The algorithm allows us to take advantage of a preconditioner that can be used to accelerate convergence.
\end{itemize}
We have implemented the LOBPCG algorithm in a nuclear structure computation software called Many-body Fermion Dynamics for nuclei (MFDn).  The implementation uses a hybrid MPI/OpenMP parallelization scheme, and has been optimized \cite{10.1007/978-3-319-46079-6_26} to achieve scalable performance for distributed-memory many-core systems such as the Cori KNL \cite{CoriKNL} system maintained at the National Energy Research Scientific Computing (NERSC) Center. 

The increased availability of high performance computing platforms equipped with general purpose GPUs has motivated us to consider modifying the many-core CPU implementation of the LOBPCG algorithm, which is written in FORTRAN 90, to enable it to run efficiently on accelerator based systems. Instead of rewriting the code using, e.g., CUDA FORTRAN~\cite{cudafortran} or OpenCL~\cite{opencl} programming models, which would take a substantial amount of work, we decided to use the OpenACC directive based programming model, with cuBLAS and cuSOLVER instead of BLAS and LAPACK, in combination with CUDA-aware MPI~\cite{MVAPICH2-GPU,cudampi}, to modify the original MPI/OpenMP based code.  In this paper, we will show how this approach can be realized. We will discuss the changes needed in order to achieve good performance on GPUs in the context of the LOBPCG algorithm. In particular, we will focus on the efficiency of the OpenACC implementation of the SpMM, which constitutes a significant portion of the overall computation, as well as the performance of the preconditioning step and other dense linear algebra operations required in the LOBPCG algorithm.  Although our discussions focus on the LOBPCG algorithm, the same techniques we developed are applicable to other block-iterative eigensolvers such as the Davidson method~\cite{Davidson1975} and the block-Lanczos algorithm~\cite{blocklanczos}.

The paper is organized as follows. In section~\ref{sec:lobpcg}, we briefly review the main components of the LOBPCG algorithm.  We discuss the hybrid MPI/OpenMP implementation of MFDn, and its key components with regard to the LOBPCG algorithm in section~\ref{sec:cpu}.  The modification of the code using OpenACC directives is presented in section~\ref{sec:gpuport}.  Numerical examples that demonstrate the performance improvement on the GPUs achieved by using OpenACC directives are presented in section~\ref{sec:examples}.  Further improvement of the existing approach is discussed in section~\ref{sec:conclude}.  We use the terminology GPU and device, as well as CPU and host, interchangeably throughout this manuscript.  Preliminary results have been presented at the 2020 OpenACC Summit \cite{oryspayev2020openacc} and at the 2021 SIAM CSE conference \cite{maris2021siamcse}.

\section{The LOBPCG Algorithm \label{sec:lobpcg}}

We denote the eigenvalues of the $n\times n$ nuclear CI Hamiltonian $H$ arranged
in an increasing order by $\lambda_1 \leq \lambda_2 \leq \cdots \leq \lambda_n$.
Their corresponding eigenvectors are denoted by $x_1$, $x_2$, ..., $x_n$.
The first $k \leq n$ eigenvectors and eigenvalues are given by
$X=[x_1,x_2,\dotsc,x_k]$ and
$\Lambda=\diag\set{\lambda_1,\lambda_2,\dotsc,\lambda_k}$, respectively,
satisfying $HX=X\Lambda$.
It is well known that $X$ is the solution to the trace minimization problem
\begin{equation}
\min_{X^TX = I} \trace(X^THX).
\label{eq:tracemin}
\end{equation}

The LOBPCG algorithm developed by Knyazev \cite{Knyazev2001} seeks to solve~\eqref{eq:tracemin}
by using the updating formula
\begin{equation}
\xcap{i+1} = \xcap{i}\ccap{i+1}_1+\wcap{i}\ccap{i+1}_2 + \pcap{i-1}\ccap{i+1}_3,
\label{eq:xupd}
\end{equation}
to approximate the eigenvector corresponding to the $k$ leftmost eigenvalues of $H$, where $\wcap{i} \in \mathbb{R}^{n\times k}$
is the preconditioned gradient of the Lagrangian
\begin{equation}
\mathcal{L}(X,\Lambda) = \frac12\trace(X^THX)
-\frac12\trace\left[(X^TX-I)\Lambda\right]
\label{eq:lag}
\end{equation}
associated with \eqref{eq:tracemin} at $\xcap{i}$, and $\pcap{i-1}$ is the search direction obtained in the $(i-1)$st iterate of the optimization procedure, and $\ccap{i+1}_1$, $\ccap{i+1}_2$, $\ccap{i+1}_3$ are a set of coefficient matrices of matching dimensions that are obtained by minimizing \eqref{eq:lag} within the subspace $\scap{i}$ spanned by 
\begin{equation}
\scap{i} \equiv \left(
\xcap{i} \ \ \wcap{i} \ \ \pcap{i-1}
\right).
\label{eq:rrsubspace}
\end{equation}

The preconditioned gradient $\wcap{i}$ can be computed as
\[
\wcap{i} = K^{-1} (H\xcap{i} - \xcap{i}\thcap{i})
\]
where $\thcap{i} = \xcapt{i}H\xcap{i}$, and $K$ is a preconditioner that approximates $H$ in some way.  The subspace minimization problem that yields the coefficient matrix $\ccap{i+1}_1$, $\ccap{i+1}_2$, $\ccap{i+1}_3$, which are three block rows of a $3k\times k$ matrix $\ccap{i+1}$, can be solved as a generalized eigenvalue problem
\begin{equation}
\left( \scapt{i} H \scap{i} \right) \ccap{i+1}
= \left( \scapt{i}  \scap{i} \right) \ccap{i+1} D^{(i+1)},
\label{eq:rreig}
\end{equation}
where $D^{(i+1)}$ is a $k\times k$ diagonal matrix containing $k$ leftmost eigenvalues of the projected matrix pencil $\left( \scapt{i} H \scap{i}, \scapt{i}  \scap{i} \right)$.
The procedure that forms the projected matrices $\scapt{i} H \scap{i}$ and $\scapt{i}  \scap{i}$ and solves the projected eigenvalue problem~\eqref{eq:rreig} is often referred to
as the \emph{Rayleigh--Ritz} procedure~\cite{parlett}.

Note that the summation of the last two terms in \eqref{eq:xupd} represents the search direction followed in the $i$th iteration, i.e.,
\begin{equation}
\pcap{i+1} = \wcap{i}\ccap{i+1}_2 + \pcap{i-1} \ccap{i+1}_3.
\label{eq:pupd}
\end{equation}

\begin{algorithm}[!tb]
\caption{The basic LOBPCG algorithm}
\label{alg:lobpcg}
\KwInput{The sparse matrix $H$, an initial guess to the $k$ desired eigenvectors $X^{(0)} \in \mathbb{R}^{n \times k}$, convergence tolerance ($tol$) and maximum number of iteration allowed ($maxiter$);}
\KwOutput{$(\Lambda,X)$, where $\Lambda$ is a $k \times k$ diagonal matrix containing the desired eigenvalues, and $X \in \mathbb{R}^{n\times k}$ contains the corresponding eignevector approximations;}

$[C^{(1)},\Theta^{(1)}]=\texttt{RayleighRitz}(H,\xcap{0})$\;
$\xcap{1} = \xcap{0} C^{(1)}$\;
$\rcap{1} = H X^{(1)} - X^{(1)} \Theta^{(1)}$\;
$P^{(0)} = \emptyset$\;
\Dor{$i=1,2,\ldots,$maxiter}
{
  $\wcap{i} = K^{-1} \rcap{i}$\;
  $\scap{i} = \left[ \xcap{i}, \wcap{i}, \pcap{i-1} \right]$\;
  $[\ccap{i+1}, \thcap{i+1}] = \texttt{RayleighRitz}(H,\scap{i})$\;
  $\xcap{i+1} = \scap{i} \ccap{i+1}$\;
  $\rcap{i+1} = H \xcap{i+1} - \xcap{i+1} \thcap{i+1}$\;
  $\pcap{i} = \wcap{i}\ccap{i+1}_2 + \pcap{i-1}\ccap{i+1}_3$\;
  Determine number of converged eigenpairs $n_c$ by checking the relative norm of each column of $\rcap{i+1}$ using the convergence tolerance $tol$\;
  exit if $n_c \geq k$\;
}
$\Lambda \leftarrow \Theta^{(i)}$; $X \leftarrow X^{(i)}$\;
\end{algorithm}

Algorithm~\ref{alg:lobpcg} outlines the main steps of the basic LOBPCG algorithm.
The most computationally costly step of Algorithm~\ref{alg:lobpcg} is the multiplication of $H$ with a set of vectors.
Although it may appear that we need to perform such calculations in steps 8 (where the projected matrix $\scapt{i} H \scap{i}$ is formed) and 10, the multiplication of 
$H$ with $\xcap{i}$, $\xcap{i+1}$ and $\pcap{i}$ can be avoided because $H\xcap{i+1}$ and $H\pcap{i}$ satisfy the following recurrence relationships
\begin{eqnarray}
 H\xcap{i+1} &=& H\xcap{i}\ccap{i+1}_1 + H\wcap{i}\ccap{i+1}_2 + H\pcap{i-1} \ccap{i+1}_3, \label{eq:upd1} \\
 H\pcap{i} &=& H\wcap{i}\ccap{i+1}_2 + H\pcap{i-1}\ccap{i+1}_3. \label{eq:upd2}
\end{eqnarray}
Therefore, the only SpMM we need to perform is $H\wcap{i}$.  For the nuclear CI calculations of interest, the dimension $n$ of the sparse symmetric matrix $H$ can be several billion, whereas $\wcap{i}$ is a tall skinny $n \times k$ matrix with $k$ typically of the order of 8 to 16.

\section{Distributed-memory Many-core CPU Implementation \label{sec:cpu}}
\subsection{Data distribution}

Because the dimension of the sparse matrix $H$ for a nuclear CI calculation can be extremely large, it is partitioned and distributed among multiple processes~\cite{sternberg2008}. Furthermore, in MFDn, we only store half of the symmetric matrix $H$ using a special data distribution scheme described below.

\begin{figure}[tb]
\centering
\includegraphics[width=0.46\textwidth]{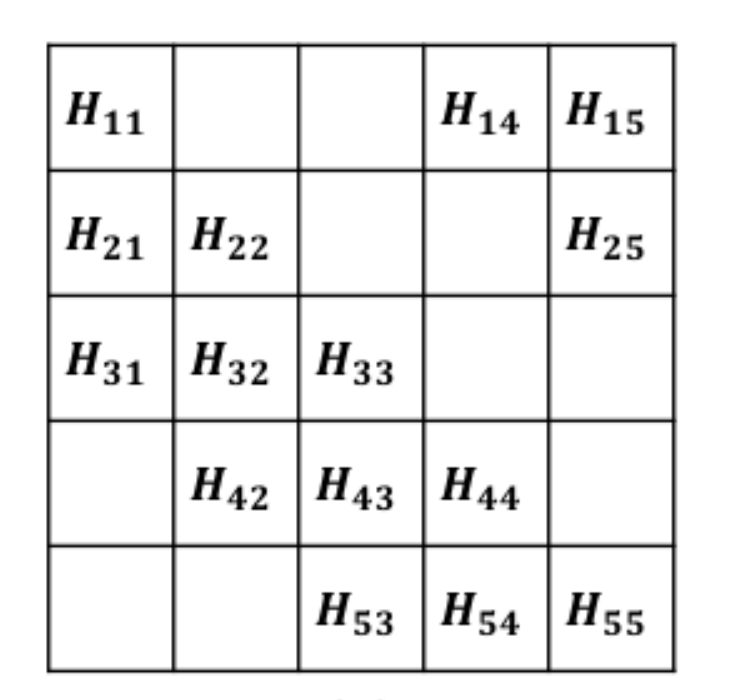}
\includegraphics[width=0.50\textwidth]{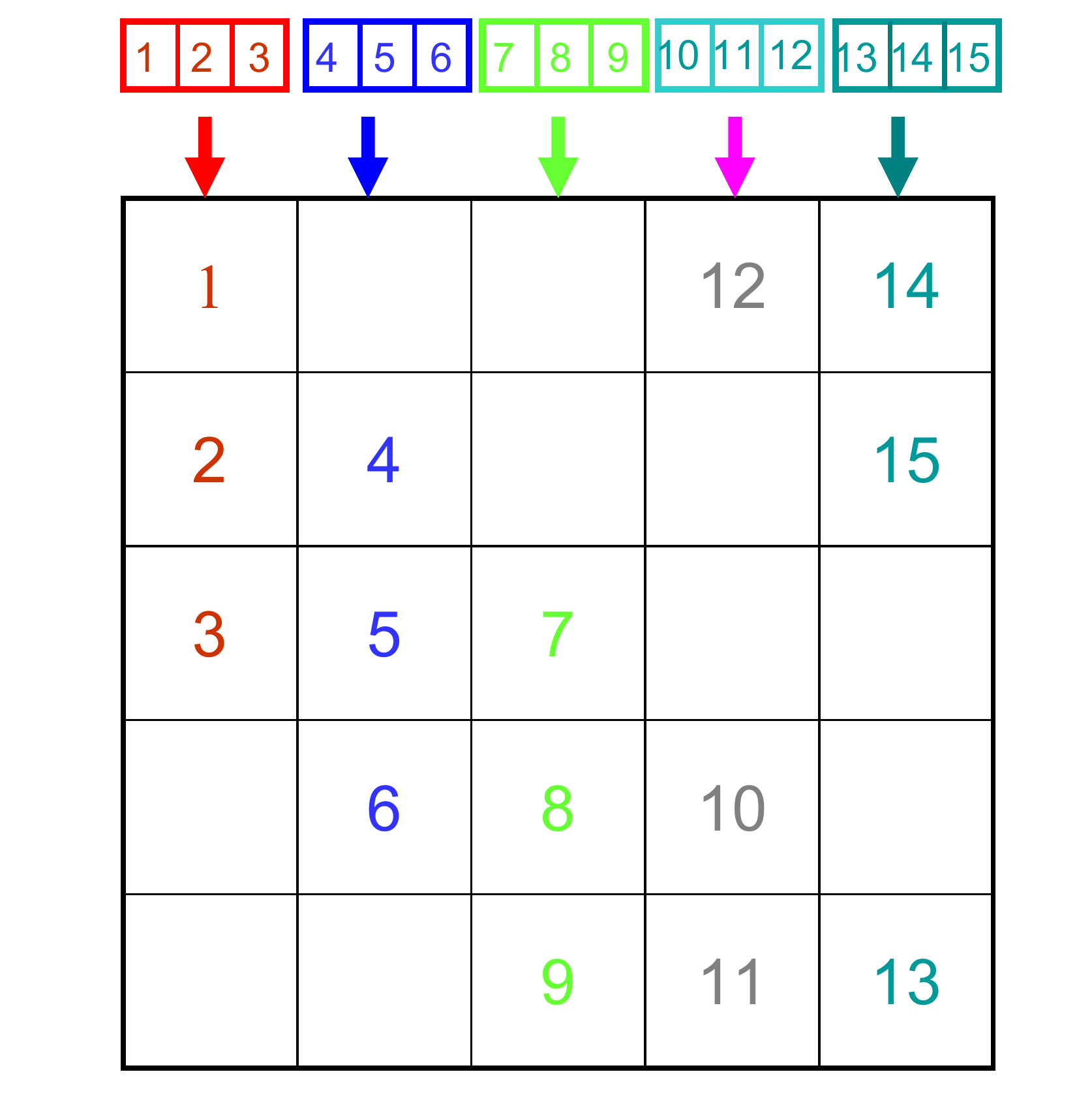}
\caption{The partition of $H$ into $5 \times 5$ sub-matrices (left) and the distribution of $H$ as well as the partition and distribution of a vector $W$ among 15 processors (right).
The integer label in the right panel represents the rank of the processes in the world communicator on which the vector segment is stored. Segments distributed among processes within the same column process group are drawn with the same color.}
\label{fig:matvecdist}
\end{figure}

We partition the rows and columns of $H$ into $n_d \times n_d$ sub-matrices, where $n_d$ is an odd integer.  We then map the $n_d (n_d+1)/2$ sub-matrices to $n_d (n_d+1)/2$ MPI processes using a column major mapping scheme developed in~\cite{topology}, see Figure~\ref{fig:matvecdist}.  These processes are grouped into (sub)communicators based on the row and column indices of the sub-matrix they hold.  There are $n_d$ row communicators as well as $n_d$ column communicators, each containing $(n_d+1)/2$ MPI ranks.
%
%

Figure~\ref{fig:matvecdist} shows how $H$ is partitioned into $5 \times 5$ sub-matrices and how the partitioned matrix is mapped to 15 processes labelled by rank 1 through 15. All sub-matrices with the same column (row) index belong to the same column (row) processor group, which is conveniently organized as a column (row) of the partitioned matrix as shown in the right panel of Figure~\ref{fig:matvecdist}. Note that the $H_{1,4}$, $H_{1,5}$ and $H_{2,5}$ sub-matrices in the upper triangular part of $H$ are the transposes of the $H_{4,1}$, $H_{5,1}$ and $H_{5,2}$ sub-matrices in the lower triangular part of $H$ respectively. Storing and working with these sub-matrices instead of their lower triangular counterpart makes the mapping of the sub-matrices to MPI processes, and the corresponding column and row communicator groups, well load-balanced.

In addition to distributing the matrix $H$, we also distribute the vectors that will be multiplied by $H$.  Each vector $x$ is first partitioned into $n_d$ sub-vectors conformal with the partitioning of $H$.  Each of these sub-vectors is then further partitioned into $(n_d+1)/2$ segments and distributed among the $(n_d+1)/2$ processes within each column group.  This partitioning of the vectors is illustrated in right panel of Figure~\ref{fig:matvecdist}. A vector is drawn on top of the matrix to illustrate how sub-vectors are aligned with the sub-matrices of $H$. Each sub-vector is drawn with a distinct color, and further partitioned into 3 segments in the column group.

\subsection{Parallel SpMM}
A customized SpMM multiplication procedure has been developed to accommodate this particular data distribution scheme in the multiplication of the distributed $H$ with a block of vectors $W$.  In the following, we denote the $i$th block of sub-vectors of $W$ by $W_i$. The distributed-memory parallel multiplication of $H$ and  $W$ is carried out in MFDn as follows:
\begin{enumerate}
\item The segments of the sub-vector $W_{i}$, which are distributed among $(n_d+1)/2$  processes within the $i$th column communicator, are gathered onto each process of that communicator using a call to {\tt MPI\_AllGatherV}.
\item The $j$th diagonal process broadcasts the gathered sub-vector $W_j=W_i$ across the $j$th row communicator in preparation for the distributed transpose SpMM computations, overlapping with the local SpMM using the local sub-matrix $H_{j,i}$, that is, $U_j = H_{j, i} W_i$. 
\item The output sub-vectors $U_j$ are reduced along the $j$th row communicator onto the $j$th diagonal process, overlapping with the local transpose SpMM on the sub-vector $W_j$, that is, $U_i = H_{i,j}W_j$.
\item The (reduced) output sub-vector $U_j$ is added to the local output sub-vector $U_i$ on the diagonal processes.
\item Finally, the sub-vectors $U_i$ are reduced and scattered into $(n_d+1)/2$ segments among the processors within the $i$th column communicator using a call to {\tt MPI\_ReduceScatter}.
\end{enumerate}

Note that we separate the multiplication of $H_{i,j}W_j$ and $H_{i,j}^T W_i$ into two separate subroutines.  This was done for two reasons: Firstly, in order to avoid race conditions (or private arrays with a reduction clause) for the OpenMP implementation of a combined local SpMM and transpose SpMM computation; and secondly, to overlap the communication along the row communicators with the local SpMM computations.  This technique was proposed in~\cite{topology} and has been shown to be very effective in both the OpenMP performance and in hiding some of the communication overhead.  We will not elaborate on this technique in this paper since our focus is on the on-device parallelization of SpMM using OpenACC.  However, in order to continue to be able to overlap communication and data transfers with local computation, we do keep this separation of the SpMM and the transpose SpMM.

\subsection{Local sparse SpMM \label{sec:local_spmm}}
On the process that holds the submatrix $H_{i,j}$, we multiply the distributed local Hamiltonian sub-matrix $H_{i,j}$ and its transpose with two blocks of sub-vectors $W_j$ and $W_i$.  A special sparse matrix storage scheme called Compressed Sparse Block with coordinate (CSB\_Coo) format~\cite{aktulga2014optimizing,aktulga2014improving} is used to make it easy to perform these two multiplications. The compressed sparse block storage is used to store the starting address of each nonzero block within $H_{i,j}$.  Each nonzero matrix element within a nonzero block of $H_{i,j}$ is represented by its local row and column indices (coordinates) as well as its numerical value. By limiting the block sizes to 32,000 (though in practice, block sizes of the order of 4,000 to 8,000 are used), we can store these local indices as 2-byte integers, and achieve almost the same memory footprint as with the more conventional Compressed Sparse Row (or Column) format for sparse matrices.

\begin{algorithm}[htbp]
\caption{Local SpMM}
\label{alg:locSpMv}
\KwInput{Matrix $H_{i,j}$ in CSB\_Coo format, a block of $n_{\rm{vec}}$ vectors $W_j$;}
\KwOutput{$U_i = H_{i,j}W_j$;}
\Dor{$i_b=1,2,\ldots, nb_{\mathrm{row}}$}
{
   \Dor{$j_b = 1, nb_{\mathrm{col}}$}
   {
      \If {$H_{i_b,j_b} \neq \emptyset$} {
          \For{\rm{each nonzero element $v$ of} $H_{i_b,j_b}$}
           {
              get the row and column indices $(i_r,i_c)$ of $v$\;
              \Dor{k = 1, $n_{\rm{vec}}$}
              {
                 $U_{i}(i_r,k) \leftarrow
              U_{i}(i_r,k) + v \cdot W_{j}(i_c,k)$\;
              }
           }
       }     
   }
}
\end{algorithm}
The multiplication of $H_{i,j}$ with a vector $W_j$ can be described by the procedure given in Algorithm~\ref{alg:locSpMv}.  Although the multiplication of $H_{i,j}^T = H_{j, i}$ with $W_i$ can be carried out in the same inner loop above, we implement the multiplication in a separate loop to allow for the local SpMM computation to be overlapped with the communication required to fetch $W_i$ as we indicated earlier.

On a multi-core processor running with multiple threads, Algorithm~\ref{alg:locSpMv} can be further parallelized by assigning each outer loop iterate (indexed by $i_b$) a single thread.   This can be done by adding an OpenMP directive immediately before the first \texttt{do} loop.  Figure~\ref{fig:spmm_omp} shows the code snippet used in MFDn for performing the local SpMM with OpenMP.  The input and output blocks of vectors are stored in the arrays \texttt{amp} and \texttt{Hamp} respectively.   For the transpose SpMM using the same data layout, the two outermost loops are simply interchanged, as well as the arguments $(r+ii)$ and $(c+ii)$ in the innermost loop, thus avoiding any race conditions, and obtaining similar performance for the SpMM and transpose SpMM.
\begin{figure}
\begin{center}
\begin{verbatim}
    !$omp parallel do default(shared) private(j, k, r, c, rbase, cbase, xcoef, ii)
    do i = 1, nrowblks
       rbase = rowCSBoffset(i) - 1
       do j = 1, ncolblks
          if (CSBnnz(i,j) > 0) then
             cbase = colCSBoffset(j) - 1
             do k = CSBnnzoffset(i,j) + 1, CSBnnzoffset(i,j) + CSBnnz(i,j)
                c = numvecs*(cbase + Hcloc(k))
                r = numvecs*(rbase + Hrloc(k))
                xcoef = Hval(k)
                !$omp simd
                do ii = 1, numvecs
                   Hamp(r+ii) = Hamp(r+ii) + xcoef * amp(c+ii)
                enddo
             enddo
          endif
       enddo
    enddo
    !$omp end parallel do
\end{verbatim}
\end{center}
\caption{Code snippet for local SpMM with OpenMP parallelization.}
\label{fig:spmm_omp}
\end{figure}

In addition to the OpenMP directive placed before the outermost loop to introduce thread-level concurrency, we also use the OpenMP SIMD directive before the innermost loop to vectorize the multiplication of $H_{i,j}$ with several vectors.  To facilitate the {\tt gather} and {\tt scatter} of these blocks of vectors among different processes within the same column communicator, multiple vectors are stored in row major order, i.e., the elements in the first row of all vectors in the block are stored contiguously, followed by the elements in the second row etc. Such a storage scheme also enhances the data locality of the SpMM.

\begin{figure}[tb]
\centering
\includegraphics[width=0.5\textwidth]{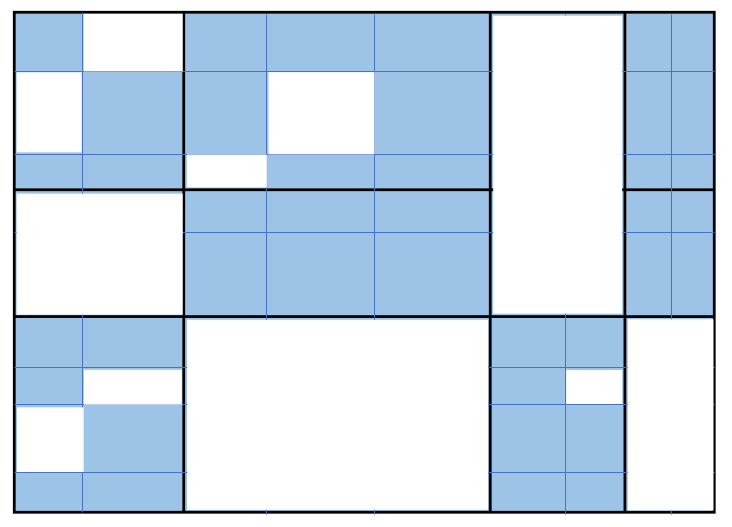}
\caption{The nonzero block and tile structure of a generic off-diagonal submatrix $H_{i,j}$. Each blue rectangle (with blue border lines) represents a nonzero tile, which is a sparse matrix. A larger rectangle with thick and black border lines represents a block. A nonzero block contains several nonzero tiles.}
\label{fig:block_tiles_demo}
\end{figure}

Finally, note that $H_{i,j}$ is a block sparse matrix, but each block itself is also sparse.  Within each nonzero block, an additional level of blocking partitions each block into zero and nonzero tiles; and at the finest level, the nonzero tiles themselves are sparse as well.
Figure~\ref{fig:block_tiles_demo} illustrates the block and tile structure of a generic off-diagonal submatrix $H_{i,j}$. Each blue rectangle (with blue border lines) represents a nonzero tile contained in a nonzero block, which is a larger rectangle with thick and black border lines.  Note that $H_{i,j}$ may contain many zero blocks that are not stored.  Within each nonzero block, the zero tiles are not stored. For a diagonal submatrix $H_{i,i}$, all diagonal blocks are nonzero.  Within a diagonal block, all diagonal tiles are nonzero (and within a diagonal tile, all diagonal matrix elements are nonzero).  These diagonal tiles are used in a preconditioner to be discussed next.

\subsection{Preconditioner}
\label{sec:precondition}
MFDn uses a block diagonal preconditioner to accelerate the convergence of LOBPCG, i.e., the preconditioner $K$ that appears in line 7 of Algorithm~\ref{alg:lobpcg} is a block diagonal matrix of the form
\begin{equation}
K = \left(
    \begin{array}{cccc}
     K_1 &     &        & \\
         & K_2 &        & \\
         &     & \ddots & \\
         &     &        & K_b
    \end{array}
    \right)\,.
\end{equation}
For $K_j$ we use the shifted diagonal tiles (introduced in the previous section) of the Hamiltonian, and $b$ is the total number of diagonal tiles. The shift used in each diagonal tile is close to the approximation to one of the desired eigenvalue.

The shifted diagonal tiles $K_j$ are distributed among all processes (not just the diagonal processes).  Each process holds several diagonal tiles.  The total number of rows in the diagonal tiles a process holds matches with the number of rows of the vector segment distributed to that process. 
We use the Formal Orthogonal Method (FOM) \cite{saadbook} to perform the preconditioning step in line 7 of Algorithm~\ref{alg:lobpcg} by solving $b$ indefinite linear systems with multiple right hand sides contained in $R^{(i)}$. 
The FOM algorithm constructs multiple Krylov subspaces associated with $K_j$ and different columns of the corresponding $R^{(i)}$ block simultaneously, from which approximations to the corresponding block of $K^{-1}R^{(i)}$ can be extracted.
Typically, three to five iterations of FOM is sufficient to achieve convergence acceleration. It is possible to use the minimal residual (MINRES)~\cite{minres} algorithms to solve these linear systems also. Since only a few iterations are performed, there is not much difference between FOM and MINRES.
Because both $K_j$ and $R^{(i)}$ are distributed on all processes, the preconditioning step is carried out on all processes with no communication.  Since each process contains a number of diagonal tiles $K_j$'s that are independent, blocks of $K^{-1}R^{(i)}$ can be computed simultaneously by multiple threads.
Within each tile, OpenMP SIMD directives are used to enable vectorization in the SpMMs within each FOM iteration. 

\subsection{Dense linear algebra operations}
\label{sec:densela}
In addition to SpMM, which constitutes the largest computational and communication work load of the LOBPCG algorithm, there are several dense linear algebra operations that need to be implemented efficiently. The Rayleigh-Ritz procedure performed in Steps 2 and 9 of Algorithm~\ref{alg:lobpcg} requires the following dense matrix-matrix multiplications
\[
G = (\xcap{0})^T A\xcap{0},
\]
and 
\begin{equation}
G^{(i)} = (\scap{i})^T A\scap{i}, \ \ O^{(i)} = (\scap{i})^T\scap{i},
\label{eq:go}
\end{equation}
to be performed, where 
$A\xcap{0}$, $A\scap{i}$ are assumed to have been computed and stored. 
In MFDn, the multiplication used to produce $G^{(i)}$ in \eqref{eq:go} is
split into six separate multiplications 
\begin{equation}
G^{(i)} = \left(
\begin{array}{ccc}
(\xcap{i})^T A\xcap{i} &  & \\
(\wcap{i})^T A\xcap{i} & (\wcap{i})^T A\wcap{i-1} & \\ 
(\pcap{i-1})^T A\xcap{i} & (\pcap{i-1})^T A\wcap{i} & (\pcap{i-1})^T A\pcap{i-1}
\end{array}
\right)
\label{eq:Glower}
\end{equation}
shown as submatrices in the lower triangular part of $G^{(i)}$ in \eqref{eq:Glower}. The subblocks in the upper triangular part of $G^{(i)}$ (which are not shown above) are simply transposes of the corresponding subblocks in the lower triangular part. A similar split is used to  obtain the overlap matrix $O^{(i)}$.

Because $\xcap{i}$, $\wcap{i}$, $\scap{i-1}$, $A\xcap{i}$, $A\wcap{i}$, $A\scap{i-1}$ are all distributed among different processes as shown in the right panel of Figure~\ref{fig:matvecdist}, the multiplications performed above are carried out locally on each process using the portion of the vectors distributed to that process. 
The local dense matrix-matrix multiplication can be carried out using optimized BLAS subroutine GEMM. 
A global reduction and broadcast (through the use of {\tt MPI\_AllReduce}) allows us to obtain and replicate $G^{(i)}$ and $O^{(i)}$ on all processes.

A similar dense matrix-matrix multiplication is performed in the 
Cholesky QR procedure used to orthonormalize columns within a matrix block $W$.  A Cholesky QR consists of the following sequence of operations.
\begin{eqnarray*}
B &=& W^T W \ \ \mbox{(Matrix-matrix multiplication)}\\
B &=& R^TR \ \ \mbox{(Cholesky factorization)}\\
W &\leftarrow & WR^{-1} \ \ \mbox{(Triangular back substitution)}
\end{eqnarray*}

We use the LAPACK library function {\tt dsygv} to solve the generalized 
eigenvalue problem \eqref{eq:rreig}, and the function {\tt dpotrf} 
to perform the Cholesky factorization of $B$. These computations can be 
replicated on all processes. The BLAS triangular solve function {\tt trsm}
is used to update $W$ locally on all processes.

The eigenvectors in $C^{(i+1)}$ obtained from Step 8 of Algorithm~\ref{alg:lobpcg} are used to update the approximate eigenvector $\xcap{i}$ and the 
search direction $\pcap{i-1}$ according to \eqref{eq:xupd} and \eqref{eq:pupd} respectively. These updates can be performed by using optimized {\tt GEMM} also. 
All these {\tt GEMM}s can be performed locally with no MPI communication because
$C^{(i+1)}$ is replicated on all processes. 

\section{GPU Implementation of LOBPCG using OpenACC \label{sec:gpuport}}

\subsection{GPU and OpenACC}

GPUs are throughput optimized processors with a Single Instruction, Multiple Thread (SIMT) programming model. These devices support massive parallelism and feature high on-device memory bandwidth making them ideal for certain data parallel workloads without complex control flow requirements. Additionally, in the low-conflict regime, performance of atomic operations is high compared to CPUs - a key feature to enable the sparse linear algebra kernels in MFDn.

MFDn is written in standard FORTRAN 90 which admits several options for a GPU implementation: CUDA FORTRAN, OpenCL, OpenMP target offload or OpenACC.  OpenACC is an attractive choice for an initial port since it is a descriptive directives based approach with stable production support in major compilers. The directive approach preserves the flexibility of running on multi-core CPUs and avoids vendor specific language lock-in. 

We use \texttt{!\$acc declare create} in combination with \texttt{!\$acc update} for all explicit data movement of (large) arrays.  
For the communication, we use CUDA-aware MPI \cite{MVAPICH2-GPU,cudampi}, which implicitly handles any necessary data movement between device and host for the MPI communication.
We adopt a descriptive approach to the use of directives for computation and exercise minimal control in order to allow the compiler and run time freedom to choose the best options. In practice, this means most loops are annotated only with \texttt{!\$acc parallel}, \texttt{!\$acc loop gang} and \texttt{!\$acc loop vector} directives with few additional clauses.

\subsection{SpMM}
We assume a one-to-one mapping of MPI processes to GPUs and that each distributed local sparse Hamiltonian $H_{i,j}$ and all other needed data fits in device  memory. CUDA-aware MPI is used to facilitate communication among devices so that we do not need to explicitly move data back and forth between hosts (where MPI communication is performed) and devices (where a majority of the computation is performed).

The easiest way to enable the local SpMM to be executed in parallel on a GPU is to replace the OpenMP parallel directives shown in the code snippets in Figure~\ref{fig:spmm_omp} with the OpenACC \texttt{parallel loop} directive for the outer loop, and the OpenACC \texttt{loop vector} directive for the inner loop, as shown in Figure~\ref{fig:spmm_acc1}.  However, this naive port does not take full advantage of the much higher levels of thread concurrency on a GPU device. On a CPU, the number of threads on a processor is often less than a few hundreds.  As a result, coarse grained concurrency achieved by decorating the outermost loop with an OpenMP parallel directive often works well. Thread overhead, which is higher on CPUs, can be minimized in this coarse grained approach. On devices, tens of thousands of lightweight threads can be simultaneously executed on multiple streaming multiprocessors (SMs) of a device. As a result, it is preferred to adopt a more fine-grained parallelization approach by generating many small tasks.  Furthermore, on GPUs, one needs a larger vector length than on current CPUs to take full advantage of the device.
\begin{figure}[tb]
\begin{center}
\begin{verbatim}
    !$acc parallel loop default(present)
    do i = 1, nrowblks
       rbase = rowCSBoffset(i) - 1
       do j = 1, ncolblks
          if (CSBnnz(i,j) > 0) then
             cbase = colCSBoffset(j) - 1
             do k = CSBnnzoffset(i,j) + 1, CSBnnzoffset(i,j) + CSBnnz(i,j)
                c = numvecs*(cbase + Hcloc(k))
                r = numvecs*(rbase + Hrloc(k))
                xcoef = Hval(k)
                !$acc loop vector
                do ii = 1, numvecs
                   Hamp(r+ii) = Hamp(r+ii) + xcoef * amp(c+ii)
                enddo
             enddo
          endif
       enddo
    enddo
    !$acc end parallel loop
\end{verbatim}
\end{center}
\caption{Code snippet of a naive OpenACC port of the local SpMM.}
\label{fig:spmm_acc1}
\end{figure}

Creating more smaller tasks can easily be realized by fusing the two outermost loops by using the OpenACC \texttt{collapse(2)} clause. In this approach, the multiplication of each nonzero block of a local Hamiltonian with a block of subvectors can be executed simultaneously. However, unlike the naive OpenACC port in which the update of the output vectors does not suffer from any write conflict, the accumulation of all products of local Hamiltonian blocks with distributed vector blocks may result in a race condition in which the same output vector block is being updated simultaneously with multiple blocked matrix-matrix products.  To avoid this race condition, an OpenACC \texttt{atomic update} clause is added around the code segment in which \texttt{Hamp} is updated. Because the overhead associated with atomic updates on a GPU device is much lower than the OpenMP atomic updates performed on a CPU, having this type of synchronization does not lead to a significant increase in the total wall clock time, as we  will show in the next section. 

In addition to fusing the two outermost loops, we also fuse the two innermost loops to expose more parallelism at the inner loop level.  The use of the OpenACC \texttt{loop vector} directive is similar in spirit to decorating the innermost loop in the OpenMP version of the local SpMM with the \texttt{simd} clause.  However, the number of vectors we work with in the inner most loop is typically 4, 8 or 16, which is significantly less than the minimum number of vector threads on current GPUs, and we cannot take advantage of all available memory banks of the device.  By vectorizing the two fused innermost loops, we expose more parallelism at the innermost loop level and create more opportunities for simultaneous access to multiple memory banks.  The atomic updates introduced to avoid potential race conditions when we fused the two outermost loops also serve to avoid potential race conditions due to fusing the two innermost loops.

The code snippet that includes the use of loop fusion, atomic updates and vectorization in the local SpMM is shown in Figure~\ref{fig:spmm_acc2}.  As we will show in the next section, these slight modifications of the original OpenACC port of the local SpMM procedure results in significant performance improvement on GPUs.

\begin{figure}[tb]
\begin{center}
\begin{verbatim}
    !$acc parallel loop collapse(2) default(present)
    do i = 1, nrowblks
       do j = 1, ncolblks
          if (CSBnnz(i,j) > 0) then
             cbase = colCSBoffset(j) - 1
             rbase = rowCSBoffset(i) - 1
             !$acc loop vector collapse(2)             
             do k = CSBnnzoffset(i,j) + 1, CSBnnzoffset(i,j) + CSBnnz(i,j)
                do ii = 1, numvecs
                   c = numvecs*(cbase + Hcloc(k)) + ii
                   r = numvecs*(rbase + Hrloc(k)) + ii
                   xcoef = Hval(k)
                   !$acc atomic update
                   Hamp(r) = Hamp(r) + xcoef * amp(c)
                   !$acc end atomic
                enddo
             enddo
          endif
       enddo
    enddo
    !$acc end parallel loop
\end{verbatim}
\end{center}
\caption{A modified OpenACC parallelization of the local SpMM that fuses two outermost loops and two inner most loops. Atomic update is used to avoid write conflicts.}
\label{fig:spmm_acc2}
\end{figure}

Further improvement in the performance of the local SpMM procedure can be made by introducing an additional level of cache blocking within each nonzero block of the local Hamiltonian. This type of blocking can further improve data locality and memory access patterns. Figure~\ref{fig:spmm_acc3} shows how this additional level of blocking is implemented. The ``\texttt{do k = ... }" loop in Figure~\ref{fig:spmm_acc2} is replaced with two nested loops in Figure~\ref{fig:spmm_acc3}.  The outer loop goes through a fixed number (defined by the variable \texttt{CacheSize} set at compile time) of nonzero matrix elements at a time.  The multiplication of these nonzero elements with the matching segment of the vector blocks are implemented in the two inner most loops that are fused and vectorized.  Also, the row and column indices as well as the nonzero elements of each chunk of size \texttt{CacheSize} are retrieved in advance and stored in thread-private arrays of length \texttt{CacheSize} so that the actual multiplication can be effectively vectorized.
\begin{figure}[tb]
\begin{center}
\begin{verbatim}
    !$acc parallel loop collapse(2) default(present)            &
    !$acc vector_length(VecLen) private(c_ar, r_ar, xcoef_ar)
    do i = 1, nrowblks
       do j = 1, ncolblks
          if (CSBnnz(i,j) > 0) then
             cbase = colCSBoffset(j) - 1
             rbase = rowCSBoffset(i) - 1
             !
             do kv = 1, CSBnnz(i,j), CacheSize
                kvmax = min(CacheSize, CSBnnz(i,j) - kv + 1)
                kvoffset = CSBnnzoffset(i,j) + kv - 1
                !$acc loop vector
                do k = 1, kvmax
                   c_ar(k) = numvecs*(cbase + Hcloc(kvoffset + k))
                   r_ar(k) = numvecs*(rbase + Hrloc(kvoffset + k))
                   xcoef_ar(k) = Hval(kvoffset + k)
                enddo
                !$acc loop vector collapse(2)
                do k = 1, kvmax
                   do ii = 1, numvecs
                      c = c_ar(k)
                      r = r_ar(k)
                      xcoef = xcoef_ar(k)
                      !$acc atomic update
                      Hamp(r+ii) = Hamp(r+ii) + xcoef * amp(c+ii)
                      !$acc end atomic
                   enddo
                enddo
             enddo
             !
          endif
       enddo
    enddo
    !$acc end parallel loop
\end{verbatim}
\end{center}
\caption{Further optimized version the OpenACC parallel local SpMM that introduces an additional level of blocking.}
\label{fig:spmm_acc3}
\end{figure}

\subsection{Preconditioning}
The parallelization of the preconditioning step of the LOBPCG algorithm on a GPU device is similar to the OpenMP parallelization of this step on a CPU.  On each device, we loop over the diagonal tiles generated on that device, and solve a linear system of equations with multiple right-hand sides by calling the subroutine \texttt{FOM\_thread}. This loop is decorated with the OpenACC \texttt{parallel loop} clause, as is shown in the code snippet in Figure~\ref{fig:precond}.  The amount of work performed in each loop iterate may be different because the size of these diagonal tiles vary from 1 to several thousand.  Therefore, the concurrency achieved in this loop is a gang level concurrency that uses a group of workers to perform each task (defined by in the subroutine \texttt{FOM\_thread}).  

\begin{figure}[!htb]
\begin{center}
\begin{verbatim}
    !$acc parallel loop default(present)                               &
    !$acc          private(indx, offset, tiledim, H, r, beta, iter)    &
    !$acc          create(W) copyin(shift)
    do indx = 1, numDiagTiles
       offset = DiagTiles_Mstate_Offset(indx) - DiagTiles_Mstate_Offset(1)
       tiledim = DiagTiles_Mstate_Offset(indx+1) - DiagTiles_Mstate_Offset(indx)
       call FOM_thread(FOMit, nact, nblk, tiledim, indx, offset,       &
            shift, W(1,1,offset+1), H, r, beta, iter)
    end do
    !$acc end parallel loop
\end{verbatim}
\end{center}
\caption{OpenACC parallelization of the preconditioner.}
\label{fig:precond}
\end{figure}

Within the subroutine \texttt{FOM\_thread}, we use OpenACC \texttt{loop vector} clauses to solve each linear system with multiple right-hand sides iteratively using many vector threads.  The main computation in each iterative FOM solver is the multiplication of a sparse matrix (diagonal tile) with a number of vectors. Because these tiles are often small, there is limited amount of concurrency we can explore in the multiplication of each (sparse) diagonal tile matrix with a block of vectors of length at most a few thousand. However, because these diagonal tiles are independent from each other, the multiplications of different diagonal tiles with different vectors can be executed simultaneously.

\subsection{Other Linear Algebra Operations}
As we pointed out in section~\ref{sec:densela}, the LOBPCG algorithm contains a number of dense linear algebra operations.  The GEMM operations required in the Rayleigh-Ritz procedure (lines 2 and 8--11 of Algorithm~\ref{alg:lobpcg} or equations \eqref{eq:xupd}) can be performed on the device by using cuBLAS GEMM calls.  In OpenACC, such calls can be made easily by simply including the {\tt cublas} module and using the same arguments in, for example, {\tt cublasdgemm}, as those used in a standard {\tt dgemm} subroutine called on a CPU.

The dense Cholesky factorization and the solution of a dense generalized symmetric eigenvalue problem can performed on the device by using cuSOLVER subroutines {\tt cuSolverDndpotrf} and {\tt cusolverDnDsygvd}. Before calling these subroutines, one has to create {\tt cuSolver} handles. A separate call to estimate buffer space required in each of these cuSOLVER subroutines needs to be made as well.  For example, the following calls 
\begin{verbatim}
    cus_status_ = cusolverDnCreate(cus_handle_)
    !$acc data copyin(pamat,eigvals)
    !$acc host_data use_device(pamat,eigvals)
    cus_status_ =  cusolverDnDsygvd_bufferSize(cus_handle_,                     &
        CUSOLVER_EIG_TYPE_1, CUSOLVER_EIG_MODE_VECTOR, CUBLAS_FILL_MODE_LOWER,  &
        width, pamat, width, pamat(1+psize), width, eigvals, liwork)
\end{verbatim}
creates a handle {\tt cus\_status\_} and returns the estimated work space requirement in {\tt liwork}.  Once the work space array {\tt rwork} is properly allocated, we make the following call
\begin{verbatim}
    !$acc data create(rwork) copy(cus_info_,pamat,eigvals)
    !$acc host_data use_device(rwork,pamat,eigvals,cus_info_)
    cus_status_ = cusolverDnDsygvd( cus_handle_,                                        &
        CUSOLVER_EIG_TYPE_1, CUSOLVER_EIG_MODE_VECTOR, CUBLAS_FILL_MODE_LOWER,          &
        width, pamat, width, pamat(1+psize), width, eigvals, rwork, liwork, cus_info_(1))
    !$acc end host_data
    !$acc end data
\end{verbatim}
to compute eigenvalues and eigenvectors of the matrices stored in {\tt pmat}.

Note that, because the cuSOLVER subroutine will be called in every iteration of the LOBPCG algorithm and the work space required in each call does not change from one iteration to another, one can place the calls to create the handle and the work space once, before the main LOBPCG iteration loop starts.  This significantly improves the performance by avoiding repeated calls to create and destroy handles, and repeated device memory allocation and deallocation at the appropriate work space size.

Finally, some of the LAPACK subroutines such as {\tt dgelqf} and {\tt dormlq} required to perform an LQ factorization of a non-square matrix $X=LQ$, where $L$ is lower triangular and $Q$ has orthonormal rows are not available in the cuSOLVER library.  To overcome this difficulty, we modified the algorithm by replacing the LQ factorization of $X$ by a QR factorization of $X^T = (LQ)^T = Q^TL^T = Q^TR$, where $R=L^T$ is upper triangular. This factorization can be performed by calling the subroutines {\tt cusolverDnDgeqrf} and {\tt cusolverDnDormqr} available in the cuSOLVER library.

\section{Performance}
\label{sec:examples}
In this section, we report the performance of the LOBPCG eigensolver in MFDn when it is executed on multiple GPUs and compare the GPU performance with the performance of the code on distributed memory many-core CPUs. Our experiments are carried out on the Cori GPU system \cite{Corigpu} at National Energy Research Scientific Computing (NERSC) Center. Each Cori GPU compute node consists of two Intel Skylake Xeon processors with 20 cores per processor, and 8 NVIDIA V100 GPUs.  Table~\ref{tbl:corigpu_spec} gives  detailed hardware and system specifications of the Cori GPU system.
\begin{table}[!ht]
\setlength\tabcolsep{4pt}
\begin{tabularx}{\textwidth}{*{3}{Z}}
\toprule
\midrule
{\bf Node CPU} & 2 x Intel Xeon (Skylake)\\
{\bf CPU Cores} & 20 @ 2.40 GHz\\
{\bf GPU} & 8 x 16 GB NVIDIA V100\\
{\bf CPU-GPU Interconnect} & PCIe 3.0\\
{\bf Interconnect} & 4 dual-port Mellanox EDR \\
\bottomrule
\end{tabularx}
\caption{System specifications of Cori GPU.}
\label{tbl:corigpu_spec}
\end{table}

We compare the performance of the OpenACC implementation of the LOBPCGsolver on the GPUs with the same implementation and same compiler on the CPUs.  Specifically, we use the NVIDIA HPC SDK 20.11 compiler for both the CPU and GPU targets, with the following compiler options
\begin{itemize}
    \item common flags: {\tt-cpp -fast -mp=numa -Mlarge$\_$arrays -Mipa -tp=skylake}
    \item flags for GPUs: {\tt-acc -ta=tesla:cc70 -Mcuda -Mcudalib=cublas,cusolver}
    \item flags for CPUs: {\tt -acc=multicore -Mmkl}
    
\end{itemize}
Note that for the GPUs we link with {\tt cuBLAS} and {\tt cuSOLVER}, whereas for the CPUs we link with the Intel Math Kernel Library MKL for the BLAS and LAPACK calls.

\subsection{Test Problems}
We use several test problems that correspond to realistic Hamiltonian matrices of different nuclei (with up to 9 protons and neutrons) represented in different configuration interaction spaces.  The dimensions of these matrices as well as the number of nonzero matrix elements (nnz) in half of each of these symmetric matrices are listed in Table~\ref{tbl:dataset}.  As we see from this table the matrix dimension of the test problems ranges from $2.9\times 10^6$ to $122.4 \times 10^6$.

\begin{table}[!ht]
\begin{center}
\begin{tabular}{|l|r|r|r|r|r|r|}
\hline
Test case & 1 & 2 & 3 & 4 & 5 & 6 \\ \hline \hline
Matrix dimension ($\times 10^6$) & $2.9$ & $10.1$ & $16.9$ & $51.9$ & $68.1$ & $122.4$ \\   \hline
\# of nonzero elements (nnz) ($\times 10^9$) & $1.1$ & $4.9$ & $13.0$ & $42.9$ & $59.1$ & $110.3$ \\  \hline
\end{tabular}
\end{center}
\caption{The dimensions of sparse matrices used in the performance test and number of nonzero matrix elements in each matrix.}
\label{tbl:dataset}
\end{table}

The smallest problem (Test 1) can fit within one GPU on a single node of Cori GPU. As the problem size becomes larger, we need to distribute the Hamiltonian matrix on multiple GPUs and perform distributed-memory parallel computation on multiple MPI ranks.  We use one GPU device per MPI process. Because each GPU on Cori GPU has 16 GB high bandwidth memory, we need to use a sufficient number of GPUs (and an appropriate number of nodes) to solve the larger problems.  Table~\ref{tbl:mpidata} shows the number of MPI processes used to solve each one of the six test problems in the following experiments.  Note that the number of nonzero matrix elements per MPI process is kept approximately fixed at about $1\times 10^9$ for each of the test problems, and thus the local compute load during the distributed SpMM should be approximately the same for each of them.  For completeness we also list the actual amount of memory required to store the distributed Hamiltonian on each GPU device or MPI process in the CSB\_Coo format described in Section~\ref{sec:local_spmm}.
\begin{table}[!ht]
\begin{center}
\begin{tabular}{|l|c|c|c|c|c|c|c|}
\hline
Test case & 1 & 2 & 3 & 4 & 5 & 6 \\ \hline \hline
\# of MPI processes & $1$ & $6$ & $15$ & $45$ & $66$ & $120$ \\  \hline
\# of nodes     & $1$ & $1$ &  $2$ &  $6$ &  $9$ &  $15$ \\  \hline
Max. \# of nnz. per MPI process ($\times 10^9$) & $1.04$ & $1.07$ & $1.03$ & $1.05$ & $0.97$ & $0.98$  \\  \hline
Max. memory per MPI process (GB) & $8.7$ & $8.7$ & $8.4$ & $8.5$ & $7.8$ & $7.8$ \\ \hline
\end{tabular}
\end{center}
\caption{The number of MPI processes and Cori GPU nodes used to solve each test problem, and the amount of memory required to hold the distributed Hamiltonian matrix per MPI process.}
\label{tbl:mpidata}
\end{table}

We compute 5 eigenpairs for each test problem and set the number of vectors in $\xcap{i}$ (and $\wcap{i}$ as well as $\pcap{i}$) to 8, which is slightly larger than the number of desired eigenpairs.  This is a common practice for improving the convergence rate of the LOBPCG algorithm as discussed in~\cite{SHAO20181}.

For each test problem, we use the same number of Cori GPU nodes (and MPI processes) to run both the CPU and GPU versions of the code, even though there are more CPU cores on each Cori GPU node than V100 GPU devices.  In the CPU runs, we use $10$ threads per MPI process, that is, we do use hyper-threading on the CPUs; the CPU performance of the OpenACC implementation is similar to that of the original OpenMP implementation.

\subsection{SpMM performance}
Since SpMM constitutes the dominant cost of the LOBPCG solver, we first examine the performance of this computational kernel.

\begin{figure}[htb]
\begin{center}
  \includegraphics[width=0.7\textwidth]{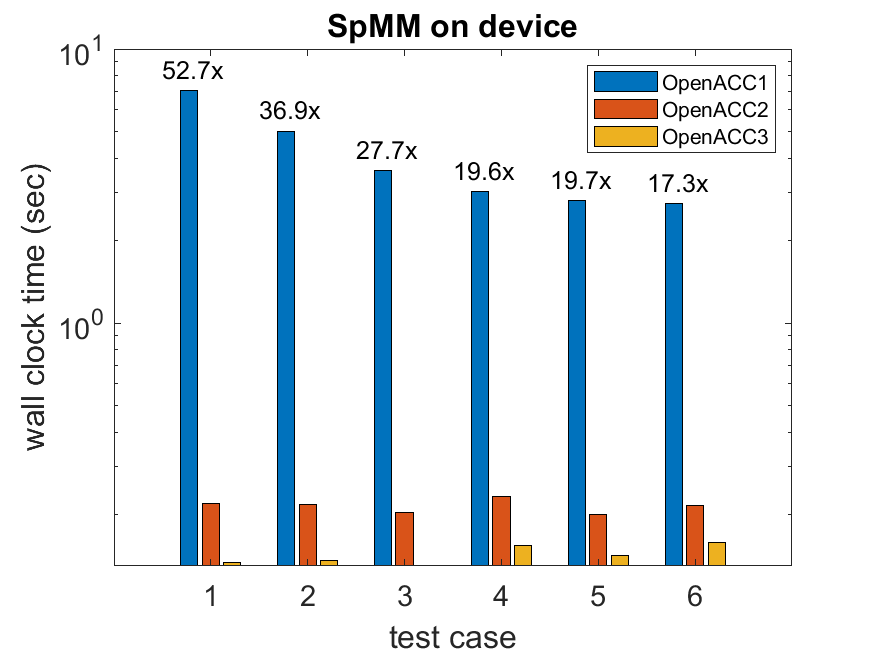}
\end{center}
\caption{A on-device performance comparison among three versions of the OpenACC implementation of the SpMM subroutine used in LOBPCG.}
\label{fig:SpMM_OpenACC1v2}
\end{figure}

In Figure~\ref{fig:SpMM_OpenACC1v2}, we first compare the three versions of the OpenACC implementation of the SpMM described in Section~\ref{sec:gpuport} on the device. 
Recall that, in the first version, which we label by OpenACC1, we simply replace the OpenMP directives in the CPU version by an OpenACC parallel loop directive on the outermost loop, and an OpenACC loop vector directive on the innermost loop, shown in Figure~\ref{fig:spmm_acc1}.  In the second version (labeled as OpenACC2), we fuse the two outer loops of the SpMM algorithm to create more tasks, and use the gang-level concurrency to launch groups of tasks.  Furthermore, we move the \texttt{vector} clause from the innermost loop to the two fused inner loops that performs a SpMM of a sparse matrix block with a block of subvectors to increase the level of vectorization and to create more fine-grained tasks, see Figure~\ref{fig:spmm_acc2}.  The \texttt{atomic update} clause is placed inside the innermost loop to ensure asynchronous updates are performed correctly.  In the third version (labeled as OpenACC3), we create a level of cache blocking within the fused inner loops to optimize memory access patterns; the size of the additional private arrays is controlled by a parameter named \texttt{CacheSize} and we also set the vector length at compile time for the vectorization of these fused innermost loops, see Figure~\ref{fig:spmm_acc3}.
The timing comparison reported in Figure~\ref{fig:SpMM_OpenACC1v2} is for time per iteration spent on local SpMM computations on the device only.  It excludes any host/device data transfer time as well as the MPI communication time.

We observe that by fusing the two outermost (row and column block) loops and moving the \texttt{vector} clause up from the innermost loop to the loop that goes through the matrix elements within each sparse block, and fuse it with the innermost loop, we can significantly improve the performance of SpMM in version 2.  Furthermore, placing the \texttt{atomic update} clause within the fused innermost loops does not appear to incur a significant overhead.  The low overhead in the OpenACC atomic update on GPUs is in sharp contrast to the high overhead in either OpenMP or OpenACC atomic update on CPUs.

Finally, by introducing an additional level of blocking within each sparse block, using the \texttt{CacheSize} parameter to control the inner block size, and tuning the vector length in that loop, we can gain additional efficiency.  We report the speedup factor defined as the ratio of OpenACC1 wall clock time over OpenACC3 wall clock time over the top of the blue bar (OpenACC1 wall clock time.) The overall  improvement in version 3 over version 1 can be as large as a speedup factor of 57 for the smallest test case and more than a speedup factor of 17 for the largest test case.

\begin{figure}[htb]
\begin{center}
     \includegraphics[width=0.495\textwidth]{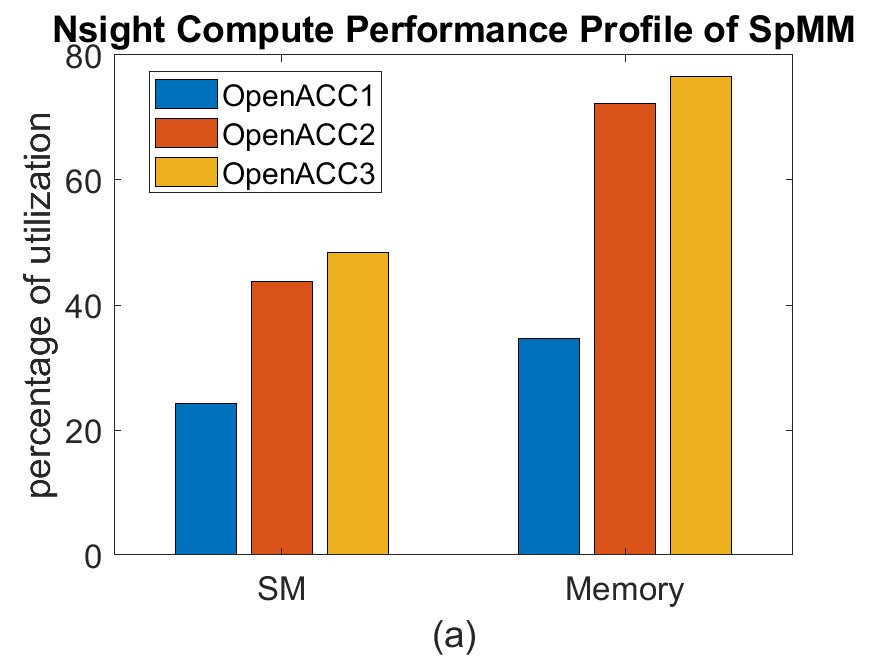}
    \includegraphics[width=0.495\textwidth]{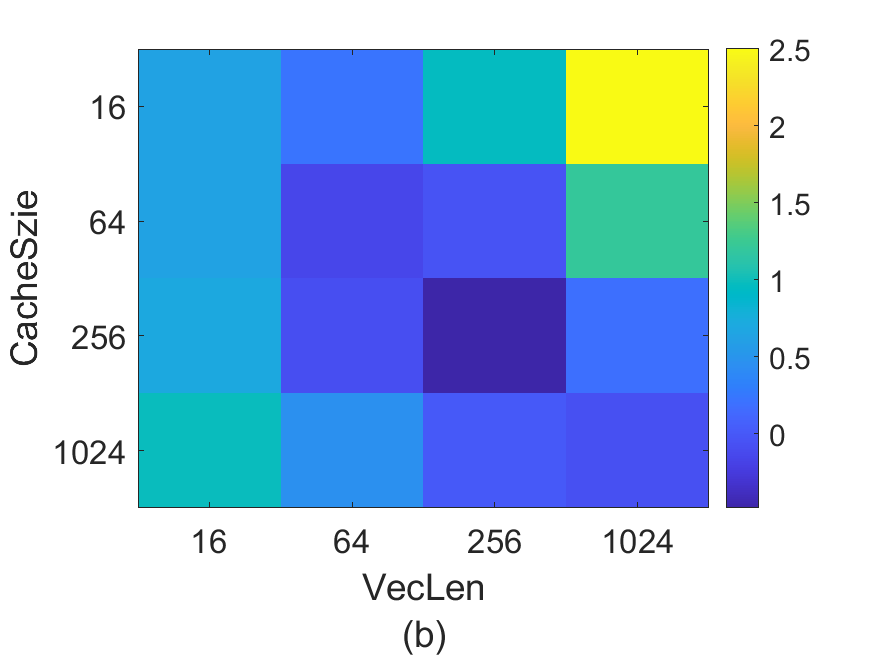}
 \caption{Left panel: A Comparison of the SM and high bandwidth memory utilization among three versions of OpenACC implementation of the local SpMM algorithm. The blue bars correspond to the naive implementation shown in Figure~\ref{fig:spmm_acc1}. The red bars correspond to the improved version shown in Figure~\ref{fig:spmm_acc2}.  The yellow bars correspond to the further improved version shown in Figure~\ref{fig:spmm_acc3}. Right panel: The variation of SpMM OpenACC3 device wallclock time with respect to vectorization length and cache size shown as a heat map for A09Z3N6\_Nmax07.}
\label{fig:SpMM_GPU_details}
\end{center}
\end{figure}

The advantage of the second and third versions of OpenACC parallelization of the SpMM algorithm for a GPU device can also be seen from a performance profile collected from the NVIDIA Nsight Compute~\cite{nsight} applied to analyze the on-device performance of the SpMM subroutine.  In Figure~\ref{fig:SpMM_GPU_details}(a), we show the GPU utilization data on one representative MPI process for test problem 3, running on 15 GPUs on two nodes.  Here we see that version 1 of the OpenACC implementation of the SpMM utilizes only $\sim$25\% of the streaming multiprocessors (SM) and $\sim$35\% of the high bandwidth memory.  The  percentages of SM and high bandwidth memory utilization increase to over 40\% and 65\% respectively after fusing the two outer loops, as well as fusing the two inner loops, as implemented in version 2 of the OpenACC implementation.  Adding an additional level of cache blocking and controlling the vector length in the inner loop, as done in version 3 of the OpenACC implementation, further increases the SM utilization to close to 50\%, and high bandwidth memory utilization to over 75\%. 

The performance of the OpenACC3 implementation is sensitive to the choice of maximum block size and the vector length specified in the OpenACC directives.  Figure~\ref{fig:SpMM_GPU_details}(b) shows variation of on device SpMM wall clock time with respect to these parameters as a heat map.  We observed that best timing in this case is achieved when both the block size and the vector length are set to 256. We use these values in all tests with OpenACC3 unless otherwise noted.

\begin{figure}[htb]
\begin{center}
  \includegraphics[width=0.495\textwidth]{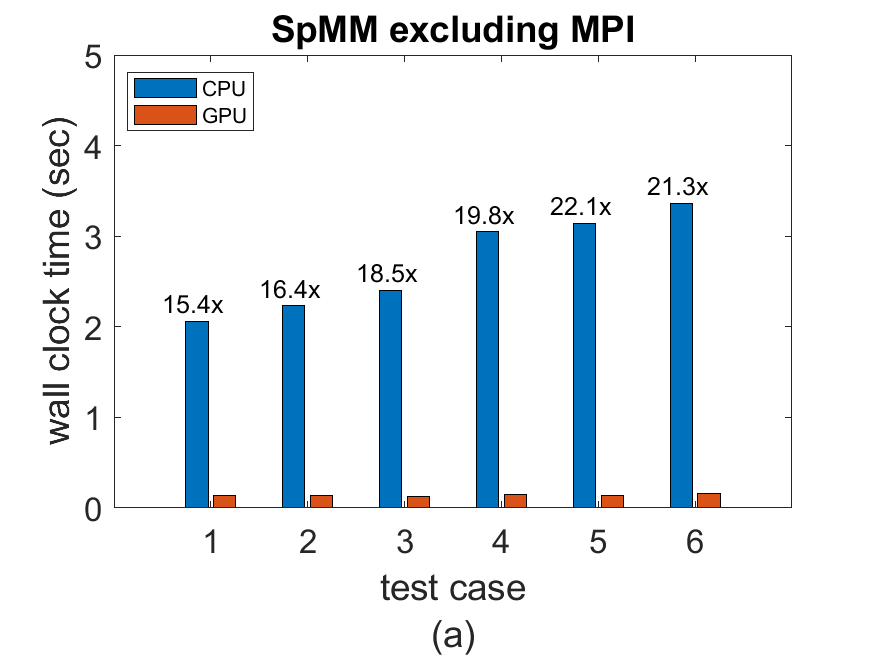}
  \includegraphics[width=0.495\textwidth]{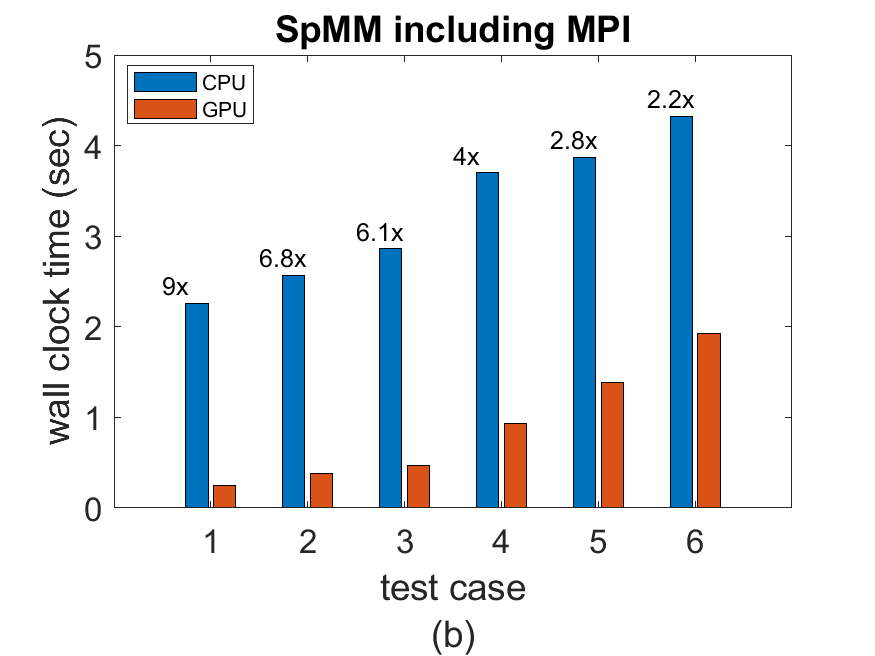}
\end{center}
\caption{Comparison between the performance of GPU/OpenACC3 implementation of the SpMM subroutine with CPU/OpenACC1 implementation (a) excluding and (b) including data transfers and MPI communication.}
\label{fig:SpMM_GPUvsCPU_noMPI}
\end{figure}

In Figure~\ref{fig:SpMM_GPUvsCPU_noMPI}, we compare the GPU performance of the OpenACC3 implementation of the SpMM with the CPU performance on the host using the OpenACC1 implementation without the need for atomics (which gives the best performance on the host, similar to the original OpenMP implementation).  In the left panel we compare the compute time only, excluding all MPI related costs.  We observe that the speedup we achieved on a GPU device ranges from about 15 for the the smallest problem to more than 21 on the larger problems.

However, when the cost associated with implicit data device-host transfers and MPI communication are included in the comparison, the speedup of the GPU version is much less 
spectacular, as can be seen from the right panel of Figure~\ref{fig:SpMM_GPUvsCPU_noMPI}.  Nonetheless we still achieve at least a factor of 2 speedup for the six test cases we are considering here.  When the larger problems are distributed over many MPI processes, communication overhead starts to dominate the SpMM runtime and cut into the on-device gains achieved by GPUs.  Indeed, including the time for data transfers and MPI communication the wall clock time for the largest of our test cases is approximately a factor of 10 larger than the local SpMM compute wall clock time on the device.


\subsection{Preconditioner performance}
\begin{figure}[htb]
\begin{center}
    \includegraphics[width=0.495\textwidth]{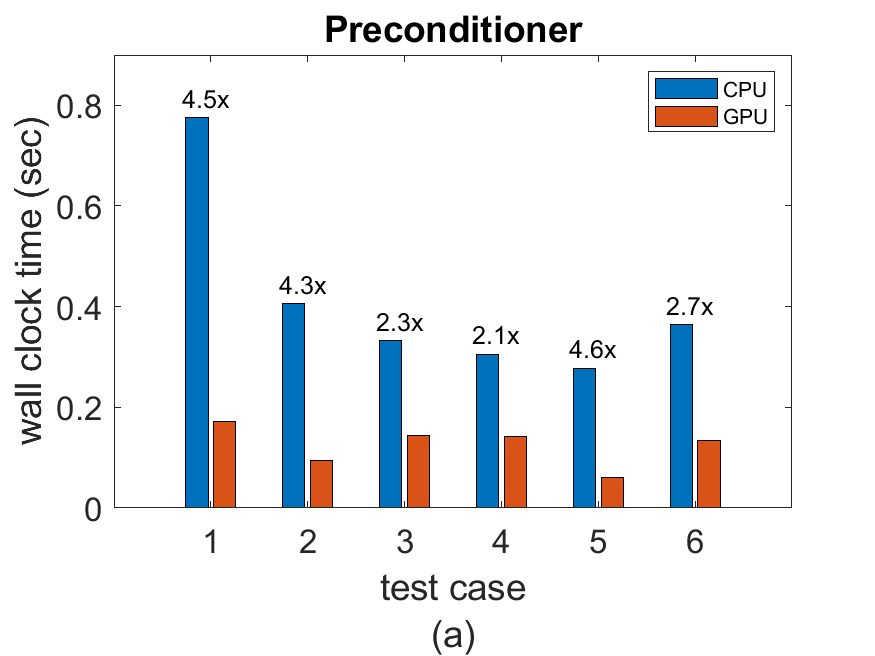}
    \includegraphics[width=0.495\textwidth]{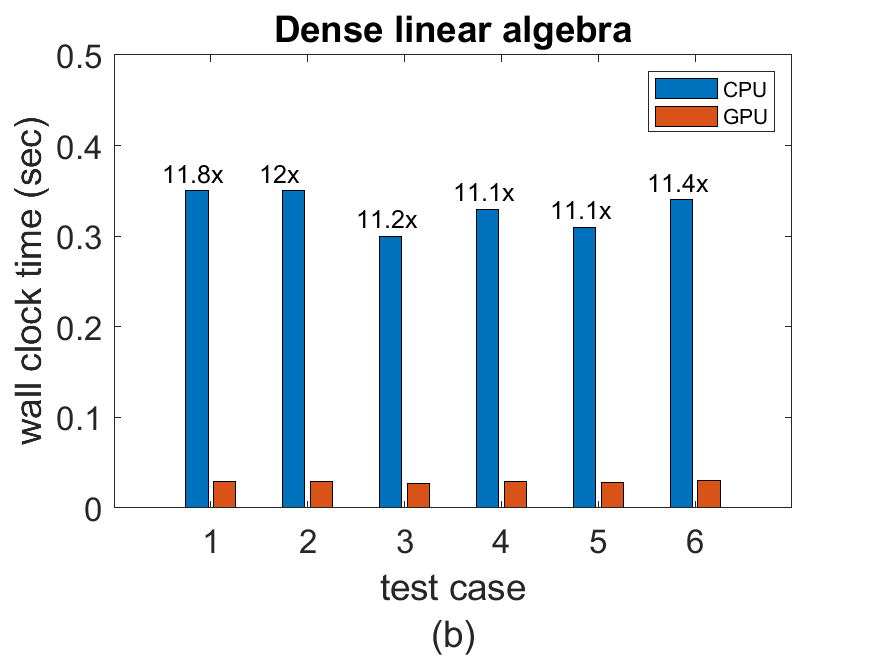}
\end{center}
\caption{Comparison between the performance of (a) the preconditioner and (b) the dense linear algebra on the GPU with that on the CPU.}
\label{fig:FOM+DLA_GPUvsCPU}
\end{figure}

\begin{figure}[htb]
\begin{center}
  \includegraphics[width=0.7\textwidth]{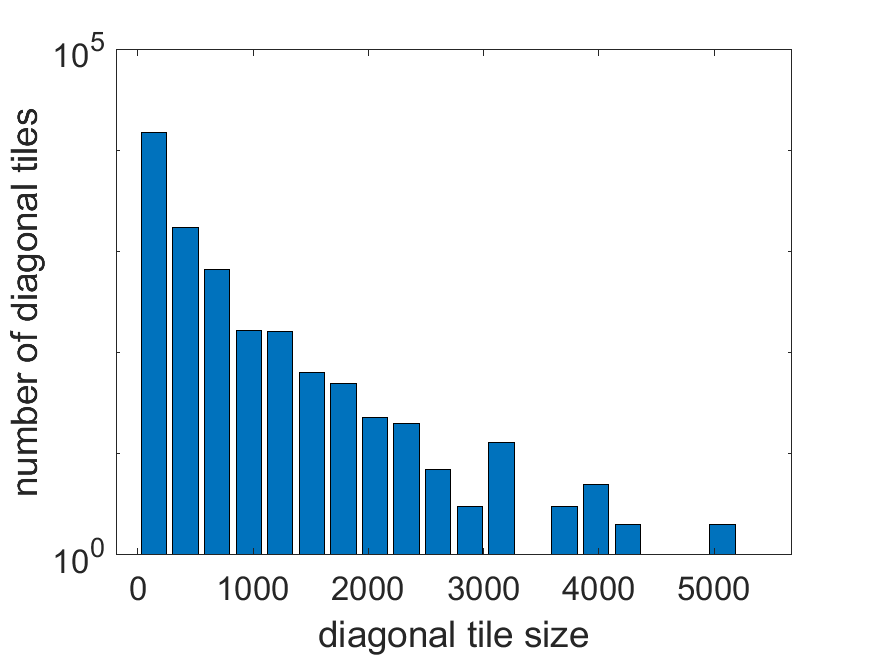}
\end{center}
\caption{A histogram of the sizes of the diagonal tiles of $H$ in test problem 1.}
\label{fig:dtile_hist}
\end{figure}

The left panel of Figure~\ref{fig:FOM+DLA_GPUvsCPU} shows how the GPU performance of the preconditioner compares with that of CPU's. As we indicated earlier, the main computation in the preconditioning step is the multiplication of a block diagonal matrix with a block of vectors. Each diagonal block is referred to as a diagonal tile. Both the diagonal tiles and vector blocks are distributed over all MPI processes. No communication is involved in this SpMM, but there is some MPI communication in the distributed dense linear algebra operations of the FOM algorithm -- this is included in the time reported in Figure~\ref{fig:FOM+DLA_GPUvsCPU}.  We observe speedup factors between 2.1 and 4.6 in the GPU version of the preconditioner.  These speedup factors are typically less than those observed for the entire Hamiltonian SpMM (except for the larger cases).  We attribute the lower speedup factors to the  relative small size of many diagonal tiles and the variation in the tile sizes. Figure~\ref{fig:dtile_hist} shows a histogram of the sizes of the diagonal tiles in the Hamiltonian of the first test problem. A majority of the tiles have sizes less than a few hundreds. The size of the largest tile is $\sim 6000$. The small sizes of many of these diagonal tiles limit the amount of concurrency OpenACC can expose in the multiplication of a single tile with a block of 8 vectors.


\subsection{Dense linear algebra performance}
The right panel of Figure~\ref{fig:FOM+DLA_GPUvsCPU} shows the speedup we achieved in the dense linear algebra operations in LOBPCG on GPU compared to CPU performance.  On GPUs, we rely on {\tt cuBLAS} and {\tt cuSOLVER} to perform these operation.  The multi-threaded MKL library is used for dense linear algebra computations on the CPUs. Overall we observe more than 11x speedup in all dense linear algebra subroutines used in the LOBPCG algorithm (excluding those used in the preconditioner).


\subsection{Overall performance}
\begin{figure}[htb]
\begin{center}
  \includegraphics[width=0.8\textwidth]{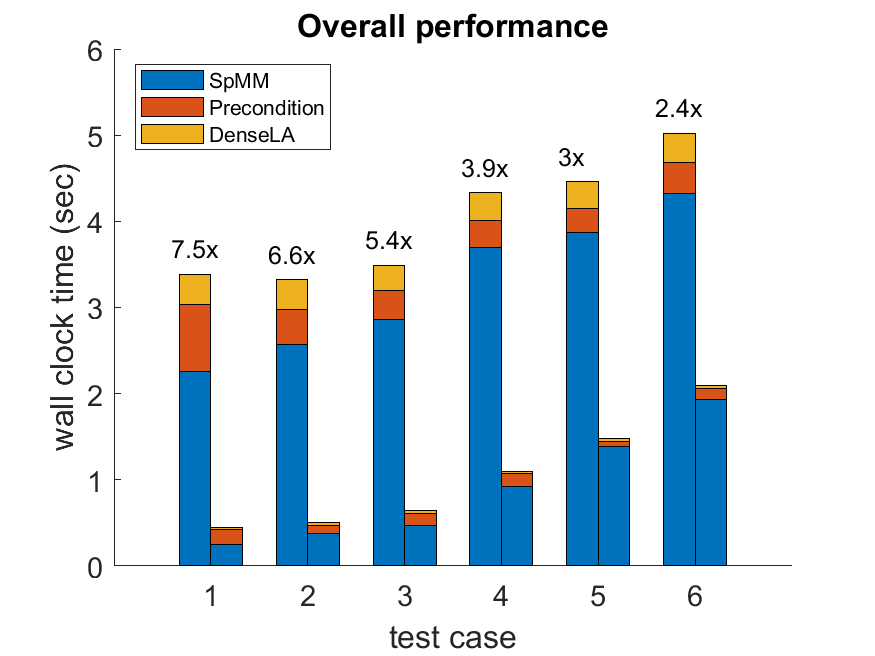}
\end{center}
\caption{Comparison between the overall performance of LOBPCG on GPU with that on the CPU.}
\label{fig:Overall_GPUvsCPU}
\end{figure}

In Figure~\ref{fig:Overall_GPUvsCPU}, we show how the overall performance of the LOBPCG eigensolver on GPUs compares with that on CPUs for all test problems, including all data transfers and MPI communication time in the iterative solver. The speedup factor we observed ranges from 2.4 for the largest problem running on 120 MPI processes to 7.5 for the smallest problem running on a single MPI process.  In addition to comparing the total wall clock time spent in each LOBPCG iteration, we also show a breakdown of timing for three main components of the LOBPCG algorithm, i.e., SpMM, preconditioning, and dense linear algebra (again, including data transfer and MPI communication time).  We observe that on GPUs, the amount of time spent in dense linear algebra is nearly negligible, whereas on CPUs, the dense linear algebra time is comparable to the preconditioner time. On both CPUs and GPUs, the wall clock time is dominated by time spent in SpMM. Therefore, the overall speedup of the LOBPCG on GPUs is limited by the speedup of the SpMM calculation.  Note that the increase of the wall clock time for the SpMM with the problem size is largely due to the increase in MPI communication and data transfer time indicating that reducing communication overhead will be crucial for large-scale computations utilizing thousands of GPUs.  On CPUs, the time spent in the preconditioner is significantly less than that spent in SpMM. However, on GPUs, the fraction of time spent in the preconditioner is comparable to that spent in the SpMM, at least for the smaller test cases, whereas for the larger test cases, running on multiple nodes, the preconditioner time represents a small percentage of the overtime wall clock time. 

\section{Conclusion \label{sec:conclude}}

In this paper, we showed how OpenACC can be used effectively to modify
a previously developed MPI/OpenMP hybrid parallel iterative LOBPCG 
eigensolver for studying nuclear structure. The OpenACC directive-based 
modification enabled the solver to run efficiently on a distributed memory 
multi-GPU system.  The use of OpenACC significantly reduces the amount 
of development effort for porting the existing eigensolver to an NVIDIA 
GPU system. However, due to the architecture difference between a general purpose GPU 
and a many-core CPU, care is needed to take advantage of 
the higher level of concurrency provided by SMs on a GPU, and high 
memory bandwidth of the device memory.  In particular, we use loop fusion
to reduce the granularity of the concurrency and increase the number of
tasks that can be executed simultaneously. Although this approach leads to 
 potential write conflicts, the extremely efficient atomic update 
in OpenACC on GPUs allows us to address this difficulty without much overhead.

We examined the parallelization of the SpMM operations which dominates 
the overall cost of the eigensolver, the preconditioning step, as well as 
some dense linear algebra operations required in the LOBPCG eigensolver.
We demonstrated that significant speedup in SpMM can be achieved on a GPU 
device by using OpenACC. However, for large problems that must be distributed
among hundreds of GPUs, the speedup factors are limited by data communication 
among different GPUs.  

We found that the speedup in the preconditioning step, which consists of performing the multiplications of many small sparse matrices with several dense vectors (viewed as a dense matrix), is also somewhat limited. This limited speedup is likely due to the small dimensions of the matrices being multiplied. In future work we will investigate better strategies to improve the performance of the preconditioning step on GPUs.

\section*{Acknowledgments}
This work was initiated during an SDSC GPU Hackathon in 2020. We would like to thank Mathew Colgrove and Brent Leback from NVIDIA, as well as Ron Caplan from Predictive Sciences Inc. for their valuable suggestions on porting the LOBPCG solver to OpenACC.
This work was supported in part by the U.\,S. Department of Energy (DOE) under grant No.~DE-SC0018223 (SciDAC\slash NUCLEI) and the National Energy Research Scientific Computing Center (NERSC) Exascale Science Applications Program (NESAP).  Computational resources were provided by NERSC, which is supported by the U.\,S. Department of Energy under Contract No.~DE-AC02-05CH11231.

\bibliographystyle{elsarticle-num}
\bibliography{refs}
\end{document}